\documentclass[aip, rsi, twocolumn]{revtex4}
\usepackage{graphicx}
\usepackage{epsfig}
\usepackage{fancyhdr}
\usepackage{hyperref}
\usepackage{mdwlist}
\usepackage{amsmath}
\usepackage{amssymb}

\begin{document}

\title{\texorpdfstring{The Twente turbulent Taylor-Couette (T$^3$C) facility: Strongly turbulent (multiphase) flow between two independently rotating cylinders}{Twente turbulent Taylor-Couette facility}}

\author{Dennis P.M. \surname{van Gils}$^1$}
\author{Gert-Wim Bruggert$^1$}
\author{Daniel P. Lathrop$^2$}
\author{Chao Sun$^1$}
\author{Detlef Lohse$^1$}
\affiliation{$^1$Department of Applied Physics and J. M. Burgers Centre for Fluid Dynamics, University of Twente, P.O. Box 217, 7500 AE Enschede, The Netherlands\\$^2$Department of Physics, IREAP and IPST, University of Maryland, College Park, Maryland 20742, USA}

\date{\today}

\begin{abstract}
\noindent
A new turbulent Taylor-Couette system consisting of two independently rotating cylinders has been constructed. The gap between the cylinders has a height of $0.927$ m, an inner radius of $0.200$ m and a variable outer radius (from $0.279$ to $0.220$ m). The maximum angular rotation rates of the inner and outer cylinder are 20 Hz and 10 Hz, respectively, resulting in Reynolds numbers up to 3.4 $\times$ 10$^6$ with water as working fluid. With this Taylor-Couette system, the parameter space ($\mathrm{Re}_i$, $\mathrm{Re}_o$, $\eta$) extends to ($2.0 \times 10^6$, $\pm1.4 \times 10^6$, 0.716 $-$ 0.909). The system is equipped with bubble injectors, temperature control, skin-friction drag sensors, and several local sensors for studying turbulent single-phase and two-phase flows. Inner-cylinder load cells detect skin-friction drag via torque measurements. The clear acrylic outer cylinder allows the dynamics of the liquid flow and the dispersed phase (bubbles, particles, fibers etc.) inside the gap to be investigated with specialized local sensors and nonintrusive optical imaging techniques. The system allows study of both Taylor-Couette flow in a high-Reynolds-number regime, and the mechanisms behind skin-friction drag alterations due to bubble injection, polymer injection, and surface hydrophobicity and roughness.
\\
\noindent [doi:\href{http://dx.doi.org/10.1063/1.3548924}{10.1063/1.3548924}]
\end{abstract}



\maketitle
\thispagestyle{fancy}

\section{Introduction}
\subsection{Taylor-Couette Flow}

Taylor-Couette (TC) flow is one of the paradigmatical systems in hydrodynamics. It consists of a fluid confined in the gap between two concentric rotating cylinders. TC flow has long been known to have a similarity \cite{bra69,dub02,eck07b} to Rayleigh-B{\'e}nard (RB) flow, which is driven by a temperature difference between a bottom and a top plate in the gravitational field of the earth \cite{ahl09,loh10}. Both of these paradigmatical hydrodynamic systems in fluid dynamics have been widely used for studying the primary instability, pattern formation, and transitions between laminar flow and turbulence \cite{cro93}. Both TC and RB flows are closed systems, i.e., there are well-defined global energy balances between input and dissipation. The amount of power injected into the  flow is directly linked to the global fluxes, i.e. angular velocity transport from the inner to the outer cylinder for the TC case, and heat transport from the hot bottom to the cold top  plate for RB case. To obtain these fluxes, one only has to measure the corresponding global quantity, namely the torque required to keep the inner cylinder rotating at constant angular velocity for the TC case, and the heat flux through the plates required to keep them at constant temperature for the RB case. In both cases the total energy dissipation rate follows from the global energy balances \cite{eck07b}. From an experimental point of view, both systems can be built with high precision, thanks to the simple geometry and the high symmetry.

For RB flow, pattern formation and flow instabilities have been studied intensively over the last century in low-Rayleigh (Ra) numbers, see e.g.\ review \cite{bod00}. With increasing Rayleigh numbers, RB flow undergoes various transitions and finally becomes turbulent. In the past twenty years, the investigation on RB flow has been extended to the high-Rayleigh-number regime, which is well beyond onset of turbulence. To vary the controlled parameters experimentally, RB apparatuses with different aspect ratios and sizes have been built in many research groups in the past twenty years \cite{cas89, geb93, cha97, gla99, nie00, bro05, sun05e, pui07, ahl09b}. Direct numerical simulation (DNS) of three-dimensional RB flow allows for quantitative comparison with experimental data up to Ra = $10^{11}$ \cite{ste10}, which is well beyond the onset of turbulence \cite{cas89}. The dependence of global and local properties on control parameters, such as Rayleigh number, Prandtl number, and aspect ratio, has been well-explored. The RB system has shown surprisingly rich phenomena in turbulent states (see, e.g., the review articles in Refs.\ \cite{ahl09} and \cite{loh10}), and it is still receiving tremendous attention (see Ref.\ \cite{NJP}).

With respect to flow instabilities, flow transitions, and pattern formation, TC flow is equally well-explored as RB flow. Indeed, TC flow also displays a surprisingly large variety of flow states just beyond the onset of instabilities \cite{pri81, and86, dom86}. The control parameters for TC flow are the inner cylinder Reynolds number $\mathrm{Re}_i$, the outer cylinder Reynolds number $\mathrm{Re}_o$, and the radius ratio of the inner to outer cylinders $\eta = r_i/r_o$. Similar to RB flow, TC flow undergoes a series of transitions from circular Couette flow to chaos and turbulence with increasing Reynolds number \cite{col65}. The flow-state dependence on the rotation frequencies of the inner and outer cylinder in TC flow at low Reynolds numbers has been theoretically, numerically and experimentally well-studied in the last century \cite{tay23, pfi81, mul87, pfi88, buc96, ess96}. This is in marked contrast to the turbulent case, for which only very few studies exist, which we will now discuss.

Direct numerical simulation of TC flow is still limited to Reynolds numbers up to $10^4$, which is still far from fully-developed turbulence \cite{don07, don08}. Experimentally, few TC systems are able to operate at high Reynolds numbers ($\mathrm{Re}_i > 10^4$), which is well beyond the onset of chaos. Smith and Townsend \cite{smi82, tow84} performed velocity-fluctuation measurements with a hot-film probe in turbulent TC flow at $\mathrm{Re}_i \sim$ 10$^4$, when only the inner cylinder was rotating. Systematic local velocity measurements on double rotating systems at high Reynolds numbers can only be found in Ref.\ \cite{wen33} from 1933. However, the measurements \cite{wen33} were performed with pitot tubes, which are an intrusive experimental technique for closed systems. Recently, Ravelet {\it et al.} \cite{rav10} built a Taylor-Couette system with independently rotating cylinders, capable of Reynolds numbers up to $10^5$, and they performed flow structure measurements in the counter-rotation region using particle image velocimetry.

The most recent turbulent TC apparatus for high Reynolds numbers ($\mathrm{Re}_i \sim 10^6$) was constructed in Texas by Lathrop, Fineberg and Swinney in 1992 \cite{lat92, lat92_phd}.  This turbulent TC setup had a stationary outer cylinder and was able to reach a Reynolds number of Re =  $1.2 \times 10^6$. Later, the system was moved to Lathrop's group in Maryland. This setup will be referred to as the Texas-Maryland TC, or T-M TC in short. The apparatus was successfully used to study the global torque, local shear stress, and liquid velocity fluctuations in turbulent states \cite{lat92, lat92a, lat92_phd, lew96_phd, lew99, ber03}.

The control parameter $\mathrm{Re}_i$ was extended to $\mathrm{Re}_i \sim 10^6$ by the T-M TC; however, the roles of the parameters $\mathrm{Re}_o$ and $\eta$ in the turbulent regime still have not been studied. Flow features inside the TC gap are highly sensitive to the relative rotation of the cylinders. The transition path from laminar flow to turbulence also strongly depends on the rotation of both the inner and the outer cylinders. A system with a rotatable outer cylinder clearly can offer much more information to better understand turbulent TC flow, and this is the aim of our setup.
On the theoretical side, Eckhardt, Grossmann, and Lohse \cite{eck07b} extend the unifying theory for scaling in
thermal convection \cite{gro00,gro01,gro02,gro04} from  RB  to TC flow,  based on the analogy between RB and TC flows. The gap ratio is one of the control parameters in TC flow, and it corresponds to the Prandtl number in Rayleigh-B{\'e}nard flow \cite{eck07b}. They define the "geometrical quasi-Prandtl number" as $\{[(1+\eta)/2]/\sqrt{\eta}\}^4$. In TC flow the Prandtl number characterizes the geometry, instead of the material properties of the liquid as in RB flow \cite{eck07b}. From considerations of bounds on solutions to the Navier-Stokes equation, Busse \cite{bus72} calculated an upper bound of the angular velocity profile in the bulk of the gap for infinite Reynolds number. The prediction suggests a strong radius ratio dependence. It is of great interest to study the role of the "geometrical quasi-Prandtl number" in TC flow. This can only be done in a TC system with variable gap ratios.

The T-M TC system was designed 20 years ago and unavoidably exhibits the limitations of its time. To extend the parameter space from only ($\mathrm{Re}_i$, 0, fixed $\eta$) to ($\mathrm{Re}_i$, $\mathrm{Re}_o$, variable $\eta$), we built a new turbulent TC system with independently rotating cylinders and variable radius ratio.

\subsection{Bubbly drag reduction}

Another motivation for building a new TC system is the increasing interest in two-phase flows, both from a fundamental and an applied point of view. For example, it has been suggested that injecting bubbles under a ship's hull will lower the skin-friction drag and thus reduce the fuel consumption; for a recent review on the subject we refer to Ref.\  \cite{cec10}. In laboratory experiments skin-friction drag reductions (DR)  by bubble injection up to $20\%$ and beyond  have been reported \cite{mad84,tak00}. However, when supplying an actual-scale ship with bubble generators, the drag reduction drops down to a few percent \cite{kod01}, not taking into account the power needed to generate the air bubbles. A solid understanding of the drag reduction mechanisms occurring in bubbly flows is still missing.

The conventional systems for studying bubbly DR are channel flows \cite{kod00}, flat plates \cite{mer89, lat03}, and cavitation flows \cite{cec10}. In these setups it is usually very difficult to control the power input into the flow, and to keep this energy contained inside the flow. In 2005, the T-M TC was outfitted with bubble injectors in order to examine bubbly DR and the effect of surface roughness \cite{ber05, ber07}. The strong point of a Taylor-Couette system, with respect to DR, is its well defined energy balance. It has been proved that the turbulent TC system is an ideal system for studying turbulent DR by means of bubble injection \cite{ber05, ber07, mur08}.

The previous bubbly DR measurements in TC flow were based only on the global torque, which is not sufficient to understand the mechanism of bubbly DR. Various fundamental issues are still unknown. How do bubbles modify the liquid flow? How do bubbles move inside the gap? How do bubbles orient and cluster? What is the effect of the bubble size? These issues cannot be addressed based on the present TC system. Ref.\ \cite{ber05} found that significant DR only appears at Reynolds numbers larger than 5 $\times$ 10$^5$ for bubbles with a radius $\sim$ 1 mm. The maximum Reynolds number of the T-M TC is around 10$^6$, which is just above that Reynolds number. A system capable of larger Reynolds numbers is therefore favored for study of this hitherto-unexplored parameter regime. The influence of coherent structures on bubbly DR can be systematically probed with a TC system when it has independently rotating cylinders.

\begin{figure*}[t]
\includegraphics[width=16cm]{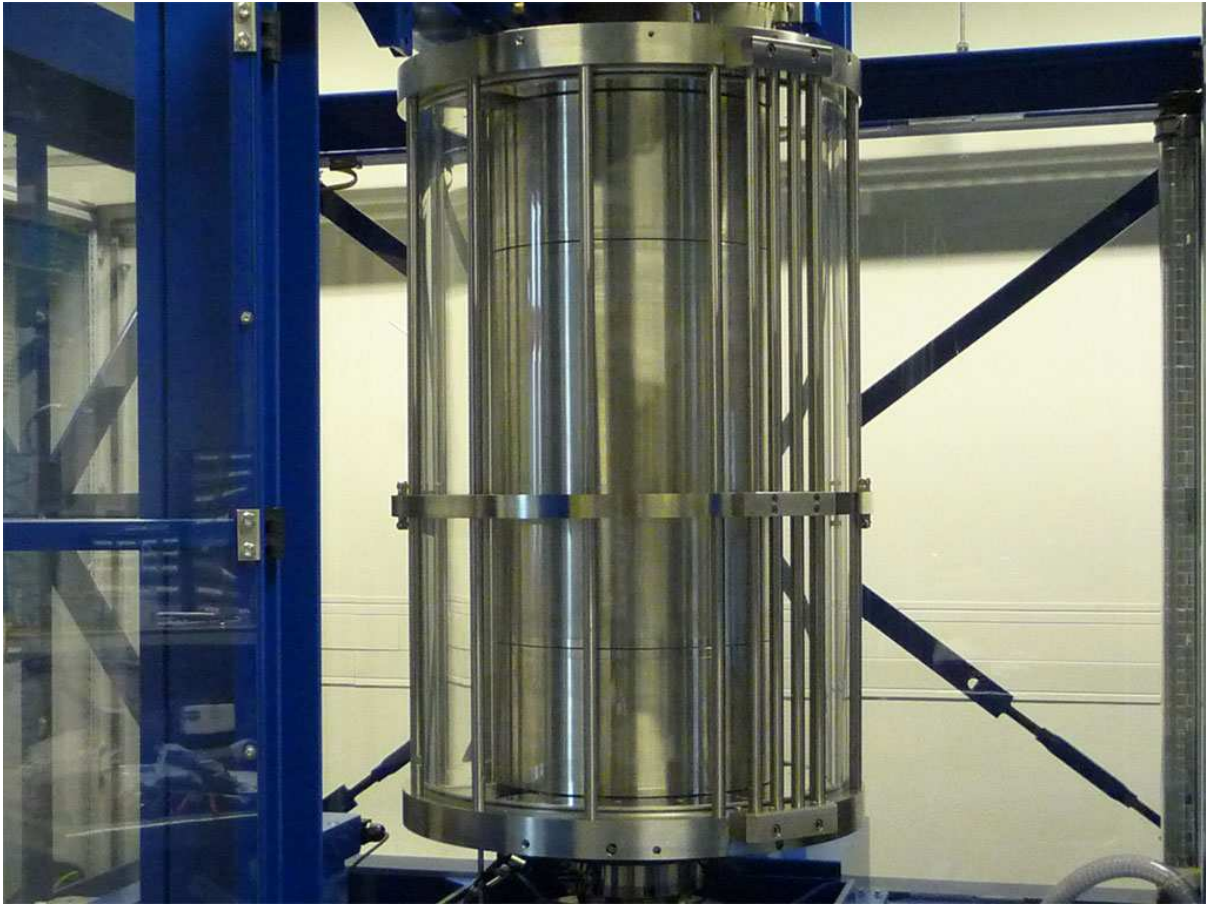}
\caption{A photograph of the T$^3$C system. The height of the cylinder is about 1 m.\label{fig:T3C_pic}}
\end{figure*}

\subsection{Twente turbulent Taylor-Couette}

Using the design of the T-M TC system as a starting point, we now present a new TC system with independently rotating cylinders, and equipped with bubble injectors, dubbed ``Twente turbulent Taylor-Couette'' (T$^3$C). We list the main features of the T$^3$C facility (the material parameters are given for water at 21$^\circ$C as the working fluid):
\begin{itemize*}
\item The inner and outer cylinder rotate independently. The maximum rotation frequencies for the inner and outer cylinder are $f_i$ = 20 Hz and $f_o$ = 10 Hz, respectively.
\item The aspect ratio and radius ratio are variable.
\item The maximum Reynolds number for the counter-rotating case at the radius ratio of $\eta = 0.716$ is 3.4 $\times$ 10$^6$. The Reynolds number for double rotating cylinders is defined as Re = $(\Omega_i r_i - \Omega_o r_o)(r_o-r_i)/\nu$, where $\Omega_i = 2\pi f_i$ and $\Omega_o = 2\pi f_o$ are the angular
velocities of the inner and outer cylinder, and $\nu$ is the kinematic viscosity of water at the operation temperature.
\item The maximum Reynolds numbers at the radius ratio of $\eta = 0.716$ are
$\mathrm{Re}_{i} = \Omega_i r_i(r_o-r_i)/\nu = 2.0 \times 10^6$ for inner cylinder rotation and
$\mathrm{Re}_{o} = \Omega_o r_o(r_o-r_i)/\nu = 1.4 \times 10^6$ for outer cylinder rotation.
\item Bubble injectors are incorporated for injecting bubbles in a range of diameters (100 $\mu m$ to 5 mm), depending on the shear strength inside the flow.
\item The outer cylinder and parts of the end plates are optically transparent, allowing for optical measurement techniques like laser Doppler anemometry (LDA), particle image velocimetry (PIV) and particle tracking velocimetry (PTV).
\item Temperature stability and rotation rate are precisely controlled.
\item Local sensors (local shear stress, temperature, phase-sensitive constant temperature anemometry (CTA), etc.) are built in or are mountable.
\end{itemize*}

Initial torque data from the T$^3$C and an upgraded T-M experiment recently appeared in Refs.\ \cite{gil10, pao10}. This paper serves to detail the T$^3$C apparatus, capabilities and measurement techniques.

\section{System description}

Figure \ref{fig:T3C_pic} is a photograph of the T$^3$C mounted in the frame. The details of the system will be described in Secs.\ \ref{sub:geom}--\ref{sub:local_sensors}.

\subsection{Geometry and materials}
\label{sub:geom}
\begin{figure}[h]
\includegraphics[width=0.3\textwidth]{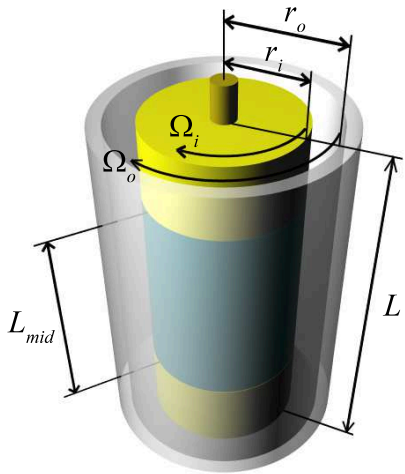}
\caption{Schematic Taylor-Couette setup consisting of two independently rotating coaxial cylinders with angular rotation rates $\Omega_i$ and $\Omega_o$. The gap between the cylinders is filled with a fluid.\label{fig:cylinders}}
\end{figure}

As shown in Fig.\ \ref{fig:cylinders}, the system contains two independently rotating cylinders of radii $r_i$ and $r_o$. The working liquid is confined in the gap between the two cylinders of width $d = r_o - r_i$. The height of the gap confined by the top and bottom plate is $L$. By design, one set of radius ratios ($\eta = r_o/r_i$) and aspect ratios $\Gamma = L/(r_o-r_i)$ of the T$^3$C nearly match those of the T-M TC in order to allow for a comparison of the results. To increase the system capacity for high Reynolds numbers, the present T$^3$C system has twice the volume compared to that of the T-M TC. The maximum Reynolds number of the T$^3$C system is 3.4 $\times$ 10$^6$ when the two cylinders at a radius ratio of $\eta = 0.716$ are counter-rotating with water at 21 $^oC$ as the working fluid. Table \ref{tb:TC_geometry} lists the geometric parameters.

\begin{table}[b]
\caption{Geometric parameters of both TC setups, with inner radius $r_i$, outer radius $r_o$, gap width $d$, radius ratio $\eta$, aspect ratio $\Gamma$ and gap volume $V_{gap}$. The outer radius of the T$^3$C gap can be varied resulting in different aspect and radius ratios. The value of 0.220 in the brackets refers to the case of a sleeve around the inner cylinder.\label{tb:TC_geometry}}
\begin{ruledtabular}
\begin{tabular}{l|l|llll}
\ & T-M TC & \multicolumn{4}{c}{T$^3$C} \\
\hline
$L$ (m)             & 0.695  & 0.927  & & & \\
$L\mathrm{_{mid}}$ (m)       & 0.406  & 0.536  & & & \\
$r_i$ (m)           & 0.160  & 0.200  & (0.220) & & \\
$r_o$ (m)           & 0.221  & 0.279  & ~0.260  & 0.240  & 0.220 \\
$d = r_o - r_i$ (m) & 0.061  & 0.079  & ~0.060  & 0.040  & 0.020 \\
$\eta = r_i / r_o$  & 0.725  & 0.716  & ~0.769  & 0.833  & 0.909 \\
$\Gamma = L / d$    & 11.43  & 11.68  & ~15.45  & 23.18  & 46.35 \\
$V_{gap}$ (m$^3$)   & 0.051  & 0.111  & ~0.080  & 0.051  & 0.024 \\
\end{tabular}
\end{ruledtabular}
\end{table}

As shown in Fig.\ \ref{fig:cylinders}, the inner cylinder (IC) consists of three separate sections IC$\mathrm{_{bot}}$, IC$\mathrm{_{mid}}$ and IC$\mathrm{_{top}}$, each able to sense the torque by means of load cell deformation embedded inside the arms connecting the IC sections to the IC drive shaft. End effects induced by the bottom and top of the TC tank are significantly reduced when focusing only on the IC$\mathrm{_{mid}}$ section. The gap between neighbouring sections is 2 mm. The material of the IC sections is stainless steel (grade 316) with a machined cylindricity (radial deviations) of better than 0.02 mm.

The outer cylinder (OC) is cast from clear acrylic, providing full optical access to the flow between the cylinders. The OC is machined to within tolerances by Blanson Ltd.\ (Leicester, UK) and the final machining was performed by Hemabo (Hengelo, Netherlands), which consisted of drilling holes in the OC for sensors, and removing the stresses in the acrylic by temperature treatment. The thickness of the OC is 25.4 mm. The bottom and top of the TC tank are connected by the OC and rotate as one piece, embedding the IC completely with the IC drive axle protruding through the top and bottom plates by means of mechanical seals.

The frame itself, as shown in blue in Fig.\ \ref{fig:T3C_pic}, is supported by adjustable air springs underneath each of its four support feet; they lift the frame fully off the ground. In combination with an inclination sensor, the frame can automatically level itself to ensure vertical alignment of the IC and OC with respect to the gravity.

\subsection{Varying gap width and IC surface properties}

Additional clear acrylic cylinders are available for this new T$^3$C system, and can be fitted between the original OC and IC. These "filler" cylinders will rotate together with the original OC and provide a way to decrease the outer radius of the gap. The nominal gap width of $0.079$ m can thus be varied to gap widths of $0.060$, $0.040$ and $0.020$ m, resulting in the aspect and radius ratios as shown in Table \ref{tb:TC_geometry}.

Instead of a smooth stainless steel IC surface, other IC surfaces with alternative chemical and/or surface structure properties can be employed, preferably in a perfectly reversible way. Thus we employed a set of cylindrical stainless steel sleeves, installed around the existing IC sections, leaving the original surface unaltered. These sleeves each clamp onto the IC sections by means of a pair of polyoxymethylene clamping rings between the sleeve and the IC. The inner radius of the gap is hence increased by 0.020 m. For good comparison at least two sets of sleeves need to be available; one set with a bare smooth stainless steel surface to determine the effect of a changing gap width, and one set with the altered surface properties. The sleeves themselves can be easily transported for surface-altering treatment.

\begin{figure*}[floatfix]
\includegraphics[height = 5 cm]{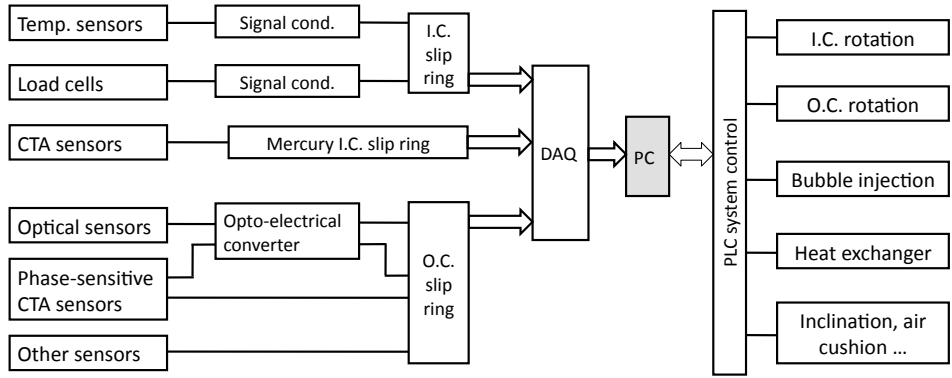}
\caption{A sketch of the signal and control system of the T$^3$C system. The abbreviations in the sketch: O.C. (outer cylinder), I.C. (inner cylinder), DAQ (data acquisition), PLC (programmable logic controllers), and PC (personal computer).  \label{fig:wiring}}
\end{figure*}

\subsection{Wiring and system control}

A sketch of the control system is shown in Fig.\ \ref{fig:wiring}. All sensors embedded inside the IC are wired through the hollow IC drive axle, and exit to a custom-made slip ring from Fabricast (El Monte, USA) on top, with the exception of the shear stress CTA (Constant Temperature Anemometry) sensors described in Sec.\ \ref{sub:local_sensors}. The slip ring consists of silver-coated electrical contact channels with 4 silver-graphite brushes per channel. The 4 brushes per channel ensure an uninterrupted signal transfer during rotation, and increase the signal-to-noise ratio. The most important signals being transferred are the signal-conditioned temperature signals and the signal conditioned load cell signals of IC$\mathrm{_{top}}$, IC$\mathrm{_{mid}}$ and IC$\mathrm{_{bot}}$, in addition to grounding and voltage feed lines. These conditioned output signals are current-driven instead of voltage-driven, leading to a superior noise suppression. From the slip ring, each electrical signal runs through a single shielded twisted-pair cable, again reducing the noise pick-up with respect to unshielded straight cables, and is acquired by data acquisition modules from Beckhoff (Verl, Germany) operating at a maximum sampling rate of 1 kHz at 16-bit resolution.

The shear stress CTA signals are fed to a liquid mercury-type slip ring from Mercotac (Carlsbad, USA). The liquid mercury inside the slip ring is used as a signal carrier and eliminates the electrical contact noise inherent to brush-type slip rings. This feature is important as additional (fluctuating) resistance between the CTA probe and the CTA controller can drastically reduce the accuracy of the measurement.

The sensors embedded in the OC are also wired through the OC drive axle and exit to a custom-made slip ring from Moog (B\"{o}blingen, Germany). Those sensors include optional CTA probes and optical fiber sensors for two-phase flow.

The T$^3$C system is controlled with a combination of programmable logic controllers (PLC) from Beckhoff which interact with peripheral electronics, and with a PC running a graphical user interface built in National Instruments LABVIEW that communicates with the PLCs. The  controlled quantities include rotation of the cylinders, bubble injection rate, temperature, and inclination angle of the system.

\subsection{Rotation rate control}

\begin{figure*}[floatfix]
\includegraphics[width=14cm]{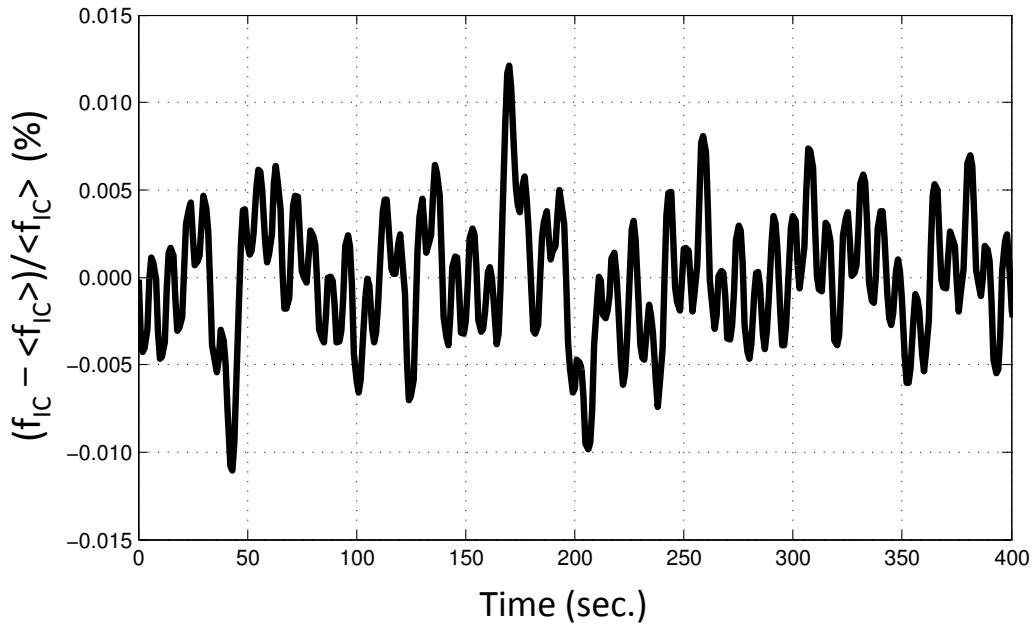}
\caption{The measured instantaneous rotation rate fluctuations normalized by the time-averaged rotation rate. Here the inner cylinder was preset to rotate at 5 Hz. \label{fig:rotation}}
\end{figure*}

The initial maximum rotation rates of the IC and OC are 20 Hz and 10 Hz, respectively. As long as the vibration velocities occurring in the main ball bearings of the T$^3$C fall below the safe threshold value of $2.8$ mm/s rms, there is room to increase these maximum rotation rates in case the total torque on the drive shaft is still below the maximum torque of 200 Nm. The measured vibration velocity in the T$^3$C system was found to be much less than that threshold, even when the system was running at these initial maximum rotation rates. The system is capable of operating at even higher rotation rates.

Both cylinders are driven by separate but identical ac motors, using a timing belt with toothed pulleys in the gear ratio 2:1 for the IC and 9:4 for the OC. Each motor, type 5RN160L04 from Rotor (Eibergen, Netherlands), is a four pole 15 kW squirrel-cage induction motor, powered by a high-frequency inverter, Leroy Somer Unidrive SP22T, operating in closed-loop vector mode. A shaft encoder mounted directly onto the ac motor shaft provides the feedback to the inverter. Electronically upstream of each inverter is a three-phase line filter and an electromagnetic compatibility filter from Schaffner (Luterbach, Switzerland), types FN3400 and FS6008-62-07, respectively. Downstream, i.e.\, between the inverter and the ac motor, are two additional filters from Schaffner, FN5020-55-34 and FN5030-55-34, which modify the usual pulse-width-modulated driving signal into a sine-wave. Important features of this filter technique are foremost the reduction of electromagnetic interference in the lab and the reduction of audible motor noise usually occurring in frequency-controlled motors.

The rotation rate of each cylinder is independently measured by magnetic angular encoders, ERM200, from Heidenhain (Schaumburg, USA), mounted onto the direct drive shafts of the IC and the OC. The magnetic line count used on the IC is 1200, resulting in an angular resolution of $0.3^\circ$, and 2048 lines, resulting in  an angular resolution of $0.18^\circ$ on the OC. The given angle resolutions do not take into account the signal interpolation performed by the Heidenhain signal controller, improving the resolution by a factor of 50 at maximum. Fig.\ \ref{fig:rotation} shows one measured time series of the rotation frequency when the system was rotating at $\langle f\mathrm{_{IC}}\rangle$ = 5 Hz. The figure shows the measured instantaneous rotation ($f\mathrm{_{IC}}$) rate fluctuation as a function of time. It is clearly shown that the rotation is stable within 0.01$\%$ of its averaged value ($\langle f\mathrm{_{IC}}\rangle$).

\subsection{Temperature control}

The amount of power dissipated by degassed water at 21 $^\circ$C, in the case of a stationary OC and the IC rotating at 20 Hz with smooth unaltered walls, is measured to be $10.0$ kW. Without cooling this would heat the $0.111$ m$^3$ of water at a rate of $\approx1.3$ K/min. As the viscosity of water lowers by $2.4\%$ per K, it is important to keep the temperature stable (to within at least $0.1$ K) to exclude viscosity fluctuations and thus errors. While the temperature-viscosity relation of water is well-tabulated and will be corrected for during measurements, other fluids like glycerin solutions might not share this feature. In the case of glycerin, a 1 K temperature increase can lower the viscosity by $\approx7\%$.

\begin{figure}[b]
\includegraphics[width=0.5\textwidth]{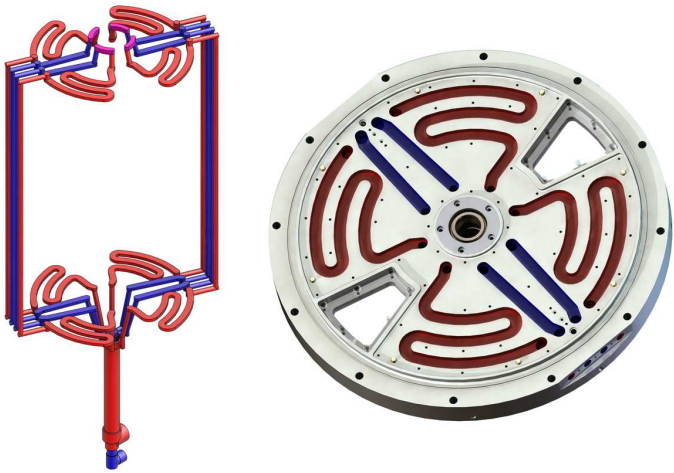}
\caption{\emph{Left}: Schematic sketch of the coolant flow through the T$^3$C. Coolant enters at the bottom rotary union (blue) and flows straight up to the top plate by piping on the outside of the TC tank. Then the coolant enters the curling channels in the top plate (red) and is fed downwards again to run through the curling channels of the bottom plate before it exits at the second rotary union. \emph{Right}: Bottom plate with the copper cover removed.\label{fig:coolant_flow}}
\end{figure}

To ensure a constant fluid temperature inside the TC tank, a 20 kW Neslab HX-750 chiller (Thermo Fisher Scientific Inc., Waltham, USA) with an air-cooled compressor and a listed temperature stability of $0.1$ K is connected to the T$^3$C. Two rotary unions, located at the bottom of the OC drive axle, are embedded inside each other and allow the coolant liquid to be passed from the stationary lab to the rotating OC, as shown in Fig.\ \ref{fig:coolant_flow}. Both the stainless steel bottom and top plate of the OC contain internal cooling channels which are covered by a 5 mm-thick nickel-coated copper plate, which in turn is in direct contact with the TC's inner volume.

\begin{figure}[t]
\includegraphics[width=0.5\textwidth]{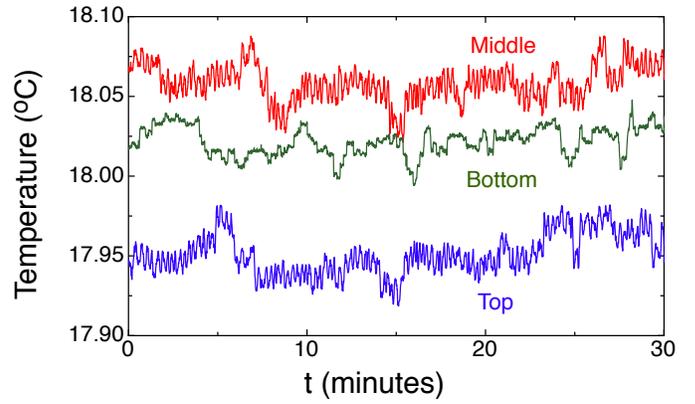}
\caption{Time traces of the measured temperature at three positions. The data were taken with the IC rotating at 20 Hz, a stationary OC and water as the working fluid. The measured power dissipation by the water was 10.0 kW.\label{fig:temp-stability}}
\end{figure}

The temperature inside the TC tank is monitored by three PT100 temperature sensors, each set up in a 4-leads configuration with pre-calibrated signal conditioners IPAQ-H$^{plus}$ from Inor (Malm\"{o}, Sweden) with an absolute temperature accuracy of $0.1$ K. The relative accuracy is better than $0.01$ K.  Each sensor is embedded at mid-height inside the wall of the hollow IC$\mathrm{_{top}}$, IC$\mathrm{_{mid}}$, and IC$\mathrm{_{bot}}$ sections, respectively (shown in Fig.\ \ref{fig:sensors}). Thus one can check for a possible axial temperature gradient across the IC. The sensors do not protrude through the wall so as to keep the outer surface smooth, leaving 1 mm of stainless steel IC wall between the sensors and the fluid. Except for the direct contact area with the wall, the PT100s are otherwise thermally isolated. Inside each IC section, the signal conditioner is mounted and its electrical wiring is fed through the hollow drive axle, ending in an electrical slip ring. The average over all three temperature sensors is used as feedback for the Neslab chiller. An example of temperature time tracers is plotted in Fig.\ \ref{fig:temp-stability}, which shows that the temperature stability is better than $0.1$ K. This data was acquired with the IC rotating at 20 Hz, a stationary OC and water as the working fluid, resulting in a measured power dissipation by the water of 10.0 kW. To check the effects of this small temperature difference on the TC flow, we calculate the Rayleigh number based on the temperature difference of $\Delta = 0.1$ K over the distance ($L_{RB}$ = 0.366 m) of the middle and top sensor positions: the result is Ra = $\beta g L_{RB}^3 \Delta/\kappa \nu$ = 5.9 $\times 10^6$ ($\beta$ is the thermal expansion coefficient, $\kappa$ the thermal diffusivity and $\nu$ the kinematic viscosity). The corresponding Reynolds number is estimated to be around $\mathrm{Re}_{RB} \sim 0.25 \times \mathrm{Ra}^{0.49}$ = 500 \cite{ahl09}, which is significantly smaller than the system Reynolds number of 2 $\times$ 10$^6$. The effects of this small temperature gradient can thus be neglected in this high-Reynolds-number turbulent flow.

%
%

\subsection{Torque sensing}

\begin{figure}[t]
\includegraphics[width=0.5\textwidth]{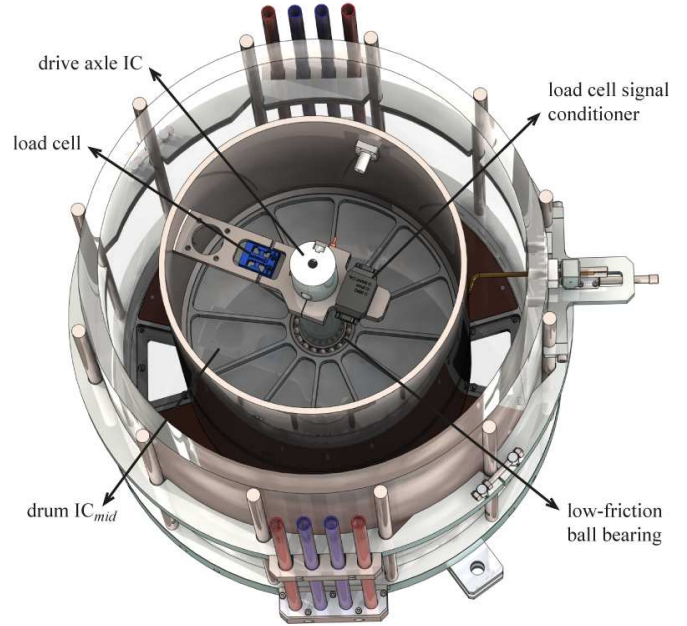}
\caption{Horizontal cut-away showing the load-cell construction inside the IC$\mathrm{_{mid}}$ drum. The load cell spans the gap in the arm, connecting the IC drive axle to the IC wall.\label{fig:IC_torque_sensor}}
\end{figure}

Each of the IC$\mathrm{_{bot}}$, IC$\mathrm{_{mid}}$ and IC$\mathrm{_{top}}$ sections are basically hollow drums. Each drum is suspended on the IC's drive axle by two low-friction ball bearings, which are sealed by rubber oil seals pressed onto the outsides of the drums, encompassing the drive axle. A metal arm, consisting of two separate parts, is rigidly clamped onto the drive axle and runs to the inner wall of the IC section. The split in the arm is bridged by a parallelogram load cell (see Fig.\ \ref{fig:IC_torque_sensor}). The load cells can be replaced by cells with different maximum-rated load capacity to increase the sensitivity to the expected torque. At this moment two different load cells, type LSM300 from Futek (Irvine, USA), are in use with a maximum-rated load capacity of 2224 N and $222.4$ N, respectively. Each load cell comes with a pre-calibrated Futek FSH01449 signal conditioner operating at 1 kHz, which is also mounted inside the drum. The electrical wiring is fed through the hollow drive axle to a slip ring on top. The hysteresis of each load cell assembly is less than 0.2 Nm, presented here as the torque equivalent. Calibration of the load cells is done by repeated measurements, in which a known series of monotonically increasing or decreasing torques is applied to the IC surface. The IC is not taken out of the frame and is calibrated \emph{in situ}. The torque is applied by strapping a belt around the IC and hanging known masses on the loose end of the belt, after having been redirected by a low-friction pulley to follow the direction of gravity.

\begin{figure}[t]
\includegraphics[height=6cm]{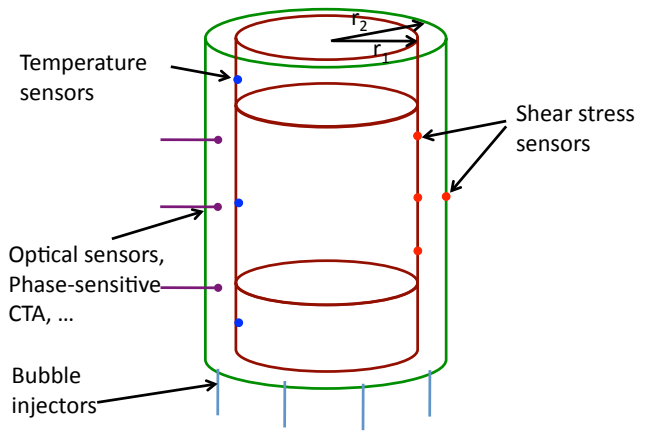}
\caption{A sketch of local sensors and bubble injectors in the system.\label{fig:sensors}}
\end{figure}

Local fluctuations in the wall-shear stress can be measured using the flush-mounted hot-film probes (type 55R46 from Dantec Dynamics) on the surfaces of the inner and outer cylinder, shown in Fig.\ \ref{fig:sensors}.

An important construction detail determines how the torque is transferred to the load cell. Only the azimuthal component is of interest and the radial and axial components, due to possible non-azimuthal imbalances inside the drum or due to the centrifugal force, should be ignored. This is accomplished by utilizing the parallelogram geometry of the load cell, lying in the horizontal plane. It is evident that during a measurement the rotation rate of the IC should be held stable to prevent the rotational inertia of the IC sections from acting as a significant extra load. The mass balance of the system is important in decreasing this effect.

\subsection{Balancing and vibrations}

All three sections of the IC and the entire OC are separately balanced, following a one-plane dynamical balancing procedure with the use of a Smart Balancer 2 from Schenck RoTec (Auburn Hills, USA). The associated accelerometer is placed on the main bottom ball bearing. The balancing procedure is reproducible to within 5 grams leading to a net vibration velocity of below 2 mm/s rms at the maximum rotation rates. According to the ISO standard 101816-1 \cite{ISO_10816_1} regarding mechanical vibrations, the T$^3$C system falls into category I, for which a vibration velocity below $2.8$ mm/s rms is considered acceptable.

Another feature is the air springs, i.e.\ pressure regulated rubber balloons, placed between the floor and each of the four support feet of the T$^3$C frame. They lift the frame fully off the ground and hence absorb vibrations leading to a lower vibration severity in the setup itself and reducing the vibrations passed on to the building.

Two permanently installed velocity transducers from Sensonics (Hertfordshire, UK), type PZDC 56E00110, placed on the top and bottom ball bearings, constantly monitor the vibration severity. An automated safety PLC circuit will stop the IC and OC rotation when tripped. Thus, dangerous situations or expensive repairs can be avoided, as this acts as a warning of imminent ball bearing failure or loss of balance.

\subsection{Bubble injection and gas concentration measurement}

Eight bubble injectors, equally distributed around the outer perimeter of the TC gap as shown in Fig.\ \ref{fig:sensors}, are built into the bottom plate of the TC tank. Each bubble injector consists of a capillary housed inside a custom-made plug ending flush with the inside wall. They can be changed to capillaries of varying inner diameter: $0.05$, $0.12$, $0.5$ and $0.8$ mm. This provides a way to indirectly control the injected bubble radius, estimated to be in the order of 0.5 mm to 5 mm, depending on the shear stress inside the TC. Smaller bubbles of radius less than 0.5 mm, or microbubbles, can be injected by replacing the capillaries inside the plugs with cylinders of porous material.

Two mass flow controllers from Bronkhorst (Ruurlo, Netherlands), series EL-Flow Select, are used in parallel for regulating the gas, i.e. filtered instrument air, with a flow rate at a pressure of 8 bars. One controller with a maximum of 36 l/min takes care of low-gas volume fractions, and the second controller with a maximum of 180 l/min of high-gas volume fractions, presumably up to 10$\%$. The gas enters the gap by a third rotary union located at the very bottom of the OC drive axle, below the coolant water rotary unions. Thus the OC drive axle has three embedded pipes running through its center, which are fed by three rotary unions at the bottom. Each pipe is split into separate channels again inside of the OC drive assembly to be routed where needed.

Vertical channels running through the near-center of the TC tank's top plate connect the tank volume to a higher located vessel in contact with the ambient air. Excess liquid or gas can escape via this route to prevent the build-up of excessive pressure. We refer to it as an expansion vessel. The expansion vessel can also be used to determine the global gas volume fraction inside the TC tank. The vessel is suspended underneath a balancer which continuously registers the vessel's mass. Starting at zero percent gas volume fraction to tare the vessel's mass, one can calculate the global gas volume fraction by transforming the liquid's mass that is subsequently pushed into the vessel by the injected gas, into its equivalent volume. In the case of a rotating OC the stationary expansion vessel can not (yet) be connected and excess liquid is collected in a stationary collecting ring encompassing the rotating top plate.

The original concept for measurement of the global gas volume fraction would not have had the restriction of a stationary OC. It makes use of a differential pressure transducer attached to the OC that measures the pressure difference between the top and bottom of the TC tank. Comparing this difference to the expected single-phase hydrostatic pressure difference, one could calculate the gas volume fraction. This method depends on the dynamic pressure being equal at the top and bottom. It fails however, due to unequal dynamic pressure induced by secondary flows. This is not unexpected as a wide variety of flow structures can exist in turbulent TC flow, like Taylor-vortices.

\subsection{Optical access and local sensors}
\label{sub:local_sensors}

Flow structure and velocity fluctuations of TC flow have been studied extensively at low Reynolds numbers, but few experiments have been performed at high Reynolds number (Re $>$ 10$^5$). Previous velocity measurements were mainly done with intrusive measurement techniques like hot-film probes. Indeed it was found that the wake effects induced by an object inside a closed rotating system can be very strong \cite{sun10}. Better velocity measurements inside the TC gap use nonintrusive optical techniques such as LDA \cite{lew96_phd}, PIV \cite{rav10}, PTV. The optical properties of the outer cylinder are hence crucial for this purpose.

The outer cylinder of the T$^3$C is transparent, and four small areas of the top and bottom plates consist of viewing portholes made of acrylic to allow for optical access in the axial direction, as can be seen in Figs.\ \ref{fig:coolant_flow} and \ref{fig:IC_torque_sensor}.
The outer cylinder was thermally treated to homogenize the refractive index and to remove stresses inside the acrylic. Thanks to this optical accessibility, all three velocity components inside the gap can be measured optically. For the velocity profile measurements, we use LDA (see Sec.\ \ref{sub:LDA}).

Various experimental studies have been done to examine bubbly DR. However, two main issues of bubbly DR in turbulent TC are still not well-studied. How do bubbles modify the turbulent flow? How do bubbles distribute and move inside the gap? It is certainly important to measure local liquid and bubble information inside the gap. Various local sensors, shown in Fig.\ \ref{fig:sensors}, can be mounted to the T$^3$C system. Here we highlight two of them: the phase-sensitive CTA and the 4-point optical fiber probe.

\subsubsection{Phase-sensitive CTA}
Optical techniques (such as LDA and PIV) are only capable of measuring flow velocities in a bubbly flow when the gas volume fraction is very low (typically less than 1$\%$). Hot-film measurements in bubbly flows also impose considerable difficulty due to the fact that liquid and gas information is present in the signal. The challenge is to distinguish and classify the signal corresponding to each phase. The hot-film probe does not provide by itself means for successful identification \cite{zen01}.  To overcome this problem, a device called phase-sensitive CTA has been developed (see Refs.\ \cite{lut05b, ber06_phd, mar10, berg10}). In this technique, an optical fiber is attached close to the hot-film so that when a bubble impinges on the sensor it also interacts with the optical fiber. The principle behind the optical fiber is that light sent into the fiber leaves the fiber tip with low reflectivity when immersed in water, and with high reflectivity when immersed in air. Hence the fiber is able to disentangle the phase information by measuring the reflected light intensity. It has proved to be an useful tool for liquid velocity fluctuation measurements in bubbly flows \cite{mar10}. Phase-sensitive CTA probes are only mounted through the holes of the outer cylinder when necessary.

\subsubsection{4-Point optical probe}
 Instead of using a single optical fiber to discriminate between phases as described in the previous paragraph, one can construct a probe consisting of four such fibers. The four fiber tips are placed in a special geometry: three fiber tips of equal length are placed parallel in a triangle, and the fourth fiber is placed in the center of gravity and protrudes past the other fiber tips (see Ref.\ \cite{gue03} for a schematic of the probe). Knowing this geometry and processing the four time series on the reflected light intensity, it becomes possible to estimate not only the size of the bubble that impinges onto the fiber tips, but also the velocity vector and the aspect ratio. To measure the bubble distribution inside the TC gap and other bubble dynamics, the 4-point optical probe is mounted through the holes of the outer cylinder only when necessary. We refer to Ref.\ \cite{xue08} for details on the measurement principle of the 4-point optical probe. Support for this probe is built into the T$^3$C by incorporating opto-electrical converters into the outer cylinder's bottom plate.

Other probes used, such as temperature and shear stress sensors will not be described as they are standard.

\section{Examples of results}
In this section we will demonstrate that the facility works by outlining our initial observations of the torque, velocity profiles and bubbly effects.

\subsection{Torque versus Reynolds number for single phase flow}

\begin{figure}[h]
\includegraphics[height=11.5cm]{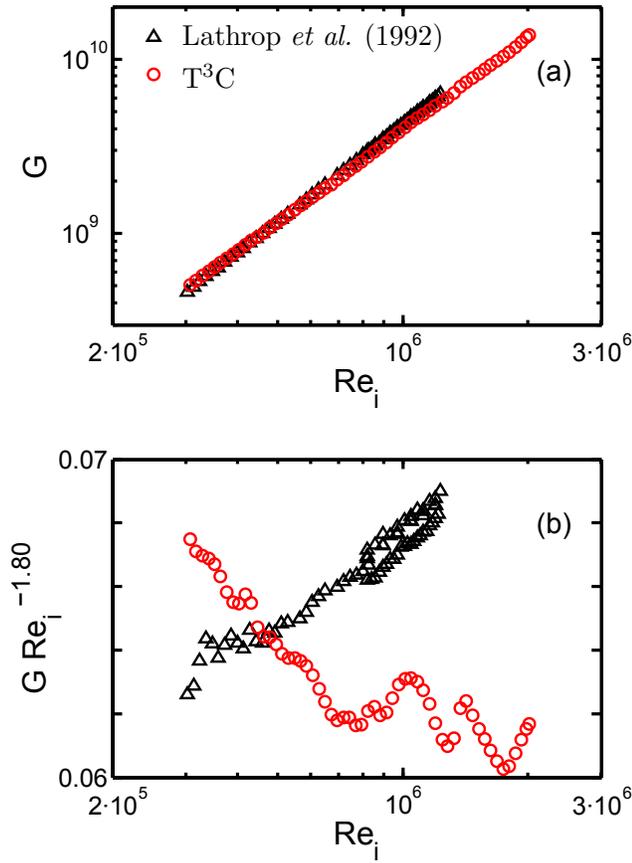}
\caption{(a) The non-dimensional torque and (b) the compensated torque $\mathrm{GRe}_i^{-1.80}$ versus Reynolds number in the high-Reynolds-number regime for the measurements with the T-M TC (open triangles) and T$^3$C (open circles) apparatuses.  \label{fig:GvsRe}}
\end{figure}

We first measure the global torque as a function of the Reynolds number in the present T$^3$C apparatus with a stationary outer cylinder. The torque ($T\mathrm{_{mid}}$) on the middle section of the inner cylinder is measured for Reynolds numbers varying from 3 $\times$ 10$^5$ to 2 $\times$ 10$^6$. We use the same normalization as Ref.\ \cite{eck07b} to define the non-dimensional torque as $\mathrm{G} = T\mathrm{_{mid}}/ 2 \pi \rho \nu^2 L\mathrm{_{mid}}$. The present measurements are performed in the T$^3$C with a radius ratio $\eta$ = 0.716, which is close to the value $\eta$ = 0.725 of the T-M TC examined by Lathrop \emph{et al.} \cite{lat92}. Figure \ref{fig:GvsRe}(a) shows G versus $\mathrm{Re}_i$ for $\mathrm{Re}_i > 3 \times 10^5$.
The fitting exponent for the data by Lathrop \emph{et al.} \cite{lat92} is 1.86, and the result of the present measurement gives 1.75.

To better compare these data, we use $\mathrm{G}^{-1.80}$ to compensate for the torque $\mathrm{GRe}_i^{-1.80}$, which is shown in Fig.\  \ref{fig:GvsRe}(b). Overall, the data of Ref.\ \cite{lat92} shows a higher exponent than the compensated value of 1.80, and the present measurement is lower.  However, both data sets clearly exhibit deviations from a single power law. The present measurement shows an oscillating trend when the Reynolds number is higher than $8 \times 10^5$. This is likely induced by transitions between different flow structures, which will be studied systematically in the T$^3$C apparatus with high scrutiny as this trend was not anticipated.

\begin{figure}[h]
\includegraphics[height=11.5cm]{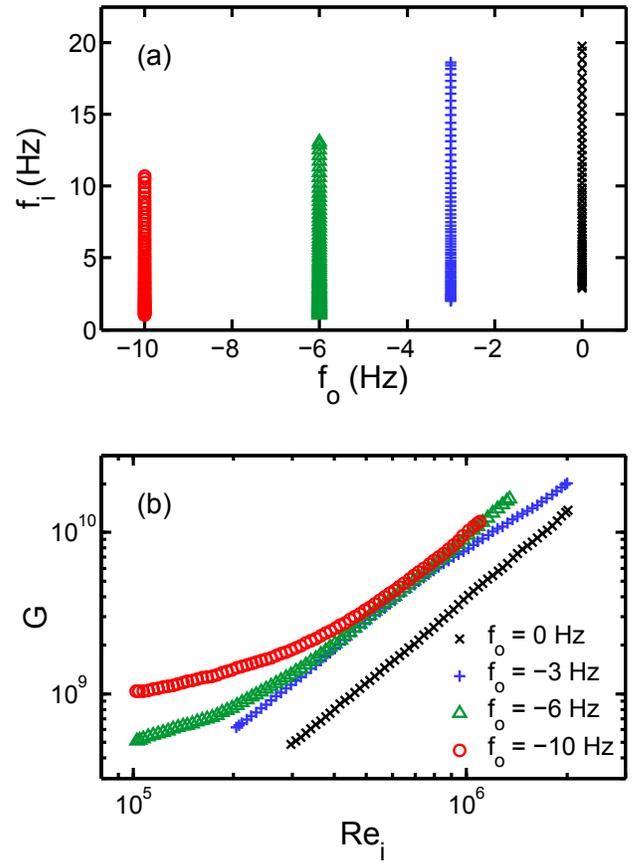}
\caption{(a) The frequency regime of counter-rotation in the present measurements. (b) The dimensionless torque G versus inner cylinder Reynolds number $\mathrm{Re}_i$  for the different outer cylinder frequencies.  \label{fig:counter}}
\end{figure}

We also examine the torque versus Reynolds number in the counter-rotation regime. The parameter space for the present measurements is shown in Fig.\ \ref{fig:counter}(a).  For three fixed outer cylinder frequencies $f_o$ = -3, -6, and -10 Hz, we measure the torque on the inner cylinder with increasing inner cylinder frequency. Fig.\ \ref{fig:counter}(b) shows the measured G versus $\mathrm{Re}_i$ for different outer cylinder frequencies. The results of pure inner rotation are also plotted in the figure for comparison. As shown in Fig.\ \ref{fig:counter}(b), both the slope and the amplitude of G versus $\mathrm{Re}_i$ for counter-rotation are different from those of pure inner rotation. It is not unexpected that the torque amplitude of the counter-rotation case exceeds the pure inner rotation case. The data for $f_o < 0$ are not simple translations of the pure inner cylinder curve, but depend on the rotation frequency of the outer cylinder in a non-trivial way. In Ref.\ \cite{gil10} we offer a unifying view to better understand $\mathrm{G}(\mathrm{Re}_i,\mathrm{Re}_o)$ in the regime of large inner and outer Reynolds numbers.

\subsection{Velocity profile measured with LDA}
\label{sub:LDA}

\begin{figure}[b]
\includegraphics[width=0.5\textwidth]{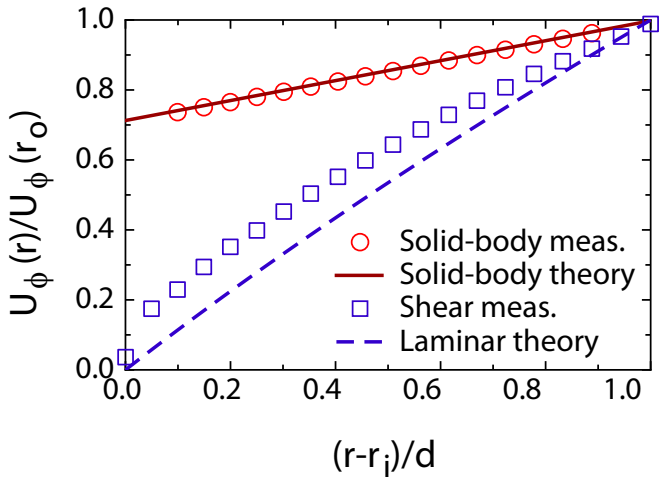}
\caption{Open circles: the azimuthal velocity profile measured with LDA when the system is in solid-body rotation  ($f_i$ = $f_o$ = 2 Hz). The solid line corresponds to $r/r_o$ of the solid-body flow. Open squares: the azimuthal velocity profile measured with LDA in the T$^3$C system with a stationary inner cylinder ($f_i$ = 0, $f_o$ = 7.2 Hz). The dashed line corresponds to the laminar profile described in Eqn. (\ref{eq:lam}), applicable to a laminar profile between the cylinders, with no top and bottom plate effects.  \label{fig:vel-profile}}
\end{figure}

Previous velocity measurements were mainly done with intrusive measurement techniques like hot-film probes. A better way for velocity measurements inside the TC gap are nonintrusive optical techniques such as LDA. Due to the curvature, the angle between two LDA beams slightly depends on the radial position along the gap. We have corrected for this via a ray-tracing calculation based on the system parameters \cite{hui10}. Refraction effects have also been taken into account. To test the reliability of the correction, we perform a velocity profile measurement when the system is in solid-body rotation, i.e. $f$ = $f_i$ = $f_o$ = 2 Hz. Open circles in Fig.\ \ref{fig:vel-profile} show the azimuthal velocity profile measured with LDA. The exact solid-body velocity profile, $r/r_o$, is shown with the solid line in the figure. The velocity has been normalized by its value on the inner wall of the outer cylinder. Figure \ref{fig:vel-profile} clearly indicates that the LDA measurements agree with the solid-body profile (within 0.6\%).

When only the outer cylinder rotates, the TC flow with an infinite aspect ratio is linearly stable. The laminar azimuthal Couette velocity profile for infinite aspect ratio (i.e. without end-plate effects) reads \cite{ll87}:
\begin{equation}
v_{lam}(r) = Ar + B/r, A = \frac{2 \pi f_o}{1-\eta^2}, B = \frac{-2 \pi f_or_i^2}{1-\eta^2}.
\label{eq:lam}
\end{equation}
The corresponding laminar profile with the present system parameters, for $f_{o}$ = 7.2 Hz, is shown with the dashed line in Fig.\ \ref{fig:vel-profile}. The open squares are the measured azimuthal velocity profile with LDA when the outer cylinder is rotating at $f_{o}$ = 7.2 Hz. The measurement is carried out with the T$^3$C at a statistically stationary state. The velocity has been normalized by its value on the inner wall of the outer cylinder. A clear deviation is found between the measurement results with the laminar profile.  The reason for this deviation is due to end-plate effects \cite{col65a,dub05,hui10} -- the laminar flow profile of type (\ref{eq:lam}) only exists when the aspect ratio of TC is infinite. More measurements are upcoming for studying velocity profiles in high-Reynolds-number Taylor-Couette flow. Certainly the end-plate effects will be significantly reduced for the turbulent cases, when the liquid velocity fluctuations dominate the flow. Additional results on velocity profiles in T$^3$C and a detailed discussion will be published elsewhere.

\subsection{Torque versus gas concentration for bubbly TC flow}

The T$^3$C apparatus was designed for studying both single- and two-phase flows. We perform torque measurements for a bubbly flow in the present T$^3$C apparatus with a stationary outer cylinder. In these first experiments with the T$^3$C facility for bubbly flow we compare the torque with the previous measurements in T-M TC by van den Berg and coworkers \cite{ber05,ber07}. The air bubbles are injected into the turbulent flow through eight capillaries located at the bottom plate. The gas fraction $\alpha$ is measured by weighing the excess volume pushed out of the system because of the added gas \cite{ber05,ber07}. The bubble radius is dictated by the strength of the shear and is typically in the range of 3$-$0.2 mm, which is kept as similar to Ref.\ \cite{ber05}. For three constant inner cylinder rotation frequencies, we measure the torque whilst slowly increasing the void fraction.

When injecting bubbles into the flow, the effective kinematic viscosity \cite{ein1906} and density are changed as follows (for low gas concentrations):

\begin{eqnarray}
\rho_\alpha = \rho (1 - \alpha),\\
\nu_\alpha = \nu (1+ \frac{5}{2}\alpha),
\end{eqnarray}
where $\alpha$ is the gas volume concentration, and $\rho$ and $\nu$ are the density and viscosity for $\alpha = 0$. The non-dimensional torque is defined based on the two-phase viscosity and density as:
\begin{equation}
\mathrm{G}(\alpha) =\frac{T\mathrm{_{mid}}}{2 \pi \rho_\alpha \nu_\alpha^2 L\mathrm{_{mid}}}=\frac{T\mathrm{_{mid}}}{2 \pi \rho(1-\alpha) \nu^2 (1+ \frac{5}{2}\alpha)^2 L\mathrm{_{mid}}},
\end{equation}
with $T\mathrm{_{mid}}$ as the measured torque on the middle section of the inner cylinder. We focus on the effect of bubbles on the torque and neglect the change of the Reynolds number.

\begin{figure}
\includegraphics[width=0.5\textwidth]{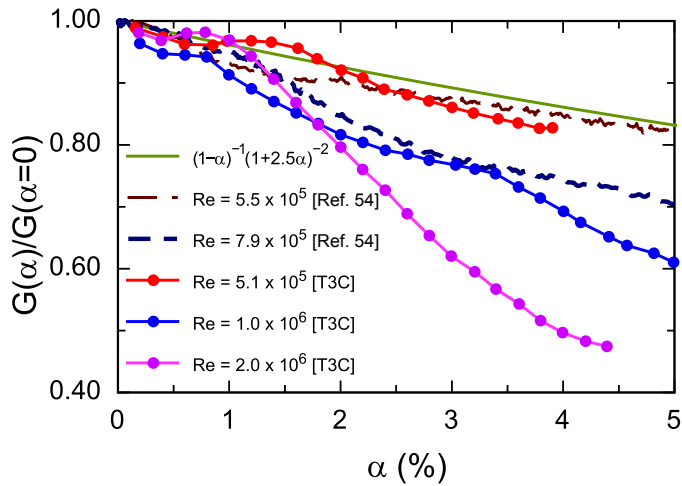}
\caption{The non-dimensional torque for different void fraction normalized by single phase torque vs.\ the gas-void fraction for different Reynolds numbers. Solid line: the effects of gas-void fraction on non-dimensional torque; dashed-lines are results measured in the T-M TC system \cite{ber07};  linked-solid dots are measurement results in the present T$^3$C system. \label{fig:bubbly_DR}}
\end{figure}

Figure \ref{fig:bubbly_DR} shows the measured torque for two-phase flow compensated for by the single-phase torque as a function of the gas volume concentration. The linked dots are the present measurement results for different Reynolds numbers. As shown in the figure, a reduction in the required torque for constant rotation rate is observed with increasing void fraction. The drag reduction is generally larger than the density and viscosity effects on the torque, as shown with the solid line in the figure. The measured maximum drag reduction increases with increasing Reynolds number, which is consistent with Ref.\ \cite{ber05}. As shown in the figure, more than 50$\%$ of drag reduction has been achieved by only adding $4\%$ gas into the system for $\mathrm{Re}_i$ = $2 \times 10^6$, which was not attainable by previous measurements in other TC apparatuses.

For quantitative comparison, the measurements in T-M TC \cite{ber07}, with Re =  $5.5 \times 10^5$ and $7.9 \times 10^5$,  are shown as the dashed lines in the figure. The agreement is clearly revealed by the nice collapse of two data sets: the data of Re = $5.1 \times 10^5$ in T$^3$C and the data of Re =  $5.5 \times 10^5$ in T-M TC. The data of Re =  $7.9 \times 10^5$ in \cite{ber07} also lies between the present data of Re =  $5.1 \times 10^5$ and  $1.0 \times 10^6$. Thus good agreement for bubbly drag reduction has been found between the T-M TC and T$^3$C systems.

Note that the new T$^3$C system shows surprisingly large drag reduction (more than 50$\%$) thanks to the system's capacity for high Reynolds numbers. This tremendous drag reduction in the high-Reynolds-number regime is very important to better understand the mechanism of turbulent drag reduction in bubbly flows, since large Reynolds numbers are more relevant to applications (e.g., carrier ships). More measurements will be done in this unexplored parameter space in the near future, and in particular we will explore the spatial bubble distribution.

\begin{figure}[h]
\includegraphics[height=5.5cm]{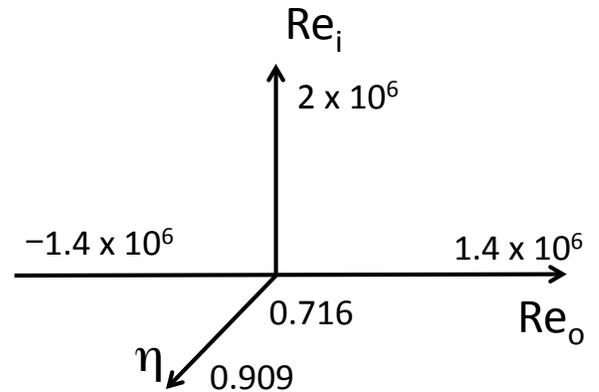}
\caption{Parameter space of the T$^3$C. \label{fig:para_space}}
\end{figure}

\section{Summary and outlook}

A new turbulent Taylor-Couette apparatus, named T$^3$C, has been developed, consisting of two independently rotating cylinders. The torque measurements for pure inner cylinder rotation agree well with previous results in the overlapped parameter regime. We also performed experiments in the unexplored parameter regime of large Reynolds number and counter-rotation of the cylinders. The data of G versus $\mathrm{Re}_i$ for the counter-rotation situation are not simple translations of the pure inner cylinder curve, but depend on the rotation frequency of the outer cylinder in a nontrivial way. The nonintrusive measurement technique LDA is applied to the system for measuring velocity profiles through the transparent outer cylinder of the T$^3$C, and it has proved to be an excellent tool for flow velocity measurements inside TC. The torque measurements of bubbly flow in the T$^3$C system showed surprisingly large drag reduction (more than 50$\%$) at high Reynolds numbers, which was not attainable by previous measurements in other TC apparatuses.

The inner cylinder Reynolds number $\mathrm{Re}_i$ of older existing turbulent TC facilities can be as high  as $10^6$. However, another two control parameters, the outer cylinder Reynolds number $\mathrm{Re}_o$ and radius ratio $\eta$, still have not been explored in the highly turbulent regime ($\mathrm{Re} \gtrsim 10^5$).
With this newly-built Taylor-Couette system, the parameter space ($\mathrm{Re}_i$, $\mathrm{Re}_o$, $\eta$) has been extended to ($2.0 \times 10^6$, $\pm1.4 \times 10^6$, 0.716 $-$ 0.909), as shown in Fig.\ \ref{fig:para_space}. Various research issues of single-phase turbulent TC flows can be studied with this new T$^3$C facility; for example, turbulent momentum transport in the unexplored parameter space of co- and counter-rotation, the role of the ``geometrical quasi-Prandtl number'' \cite{eck07b} on turbulent momentum transport, angular momentum/velocity profiles, and the connections with the global torque, and statistics of turbulent fluctuations of different velocity components. These studies will bridge the gap between RB flow and TC flow in the turbulent regime toward a better understanding of closed turbulent systems.

\begin{figure}[h]
\includegraphics[width=0.5\textwidth]{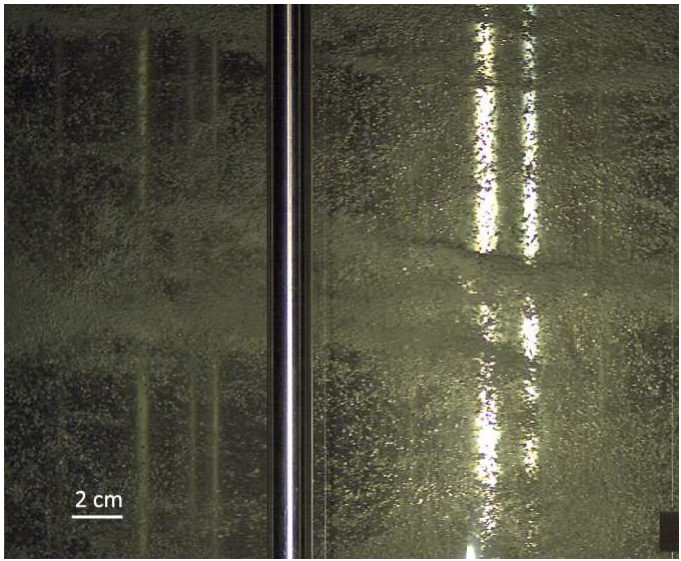}
\caption{A snapshot of the bubble distribution in the T$^3$C, $\mathrm{Re}_i = 1.0 \times 10^6$. The outer cylinder is stationary. \label{fig:bubble-dis}}
\end{figure}

The injection of bubbles offers another degree in the parameter space for studying two-phase TC flows. The study of the bubbly DR can be improved further by combining the global torque and the local measurements of the bubble velocity and distribution (see Fig.\ \ref{fig:bubble-dis}). Apart from studying bubbly DR, T$^3$C is also an ideal system for studying other issues in dispersed multiphase flows; for example, the Lagrangian properties of particles/bubbles in turbulence \cite{loh08, tos09, mar10}, and liquid velocity fluctuations induced by dispersed bubbles/particles \cite{lan91, ren05}. It was found that even in homogeneous and isotropic turbulence, particles, drops and bubbles are not distributed homogeneously, but cluster \cite{cal08b}. One example of bubble clustering in turbulent TC flow is shown in Fig.\ \ref{fig:bubble-dis}. Given the rich flow structures inside turbulent TC flow, it is of great interest to quantify particles/bubbles clustering with varying turbulent structures, and this will also be done in future research.

\begin{acknowledgments}

This study was financially supported by the Technology Foundation STW of the Netherlands.  The authors would like to thank S.G. Huisman for velocity profile measurements, G. Ahlers, R. Delfos, S. Grossmann, T. Mullin, D. Narezo, G.P. Pfister, E. van Rietbergen, R.J.A.M. Stevens, T. van Terwisga, and J. Westerweel  for scientific discussions. We also gratefully acknowledge technical contributions by B. Benschop, S.J. Boorsma, M. Bos, G. Mentink, R. Nauta, R.O. Heuvel, S. Bekmann, and J. Schepers. D.P. Lathrop was supported by a grant from the NSF/DMR of the USA.

\end{acknowledgments}




\end{document}